\documentclass{emulateapj}

\usepackage{verbatim,graphicx,epsfig,dcolumn,color}
\definecolor{red}{rgb}{0.75,0.0,0.0}
\definecolor{green}{rgb}{0.0,0.75,0.0}
\definecolor{blue}{rgb}{0.0,0.0,0.75}

\slugcomment{Draft version \today}

\shorttitle{Pulsation-triggered mass loss from AGB stars}
\shortauthors{McDonald et al.}

\begin{document}

\title{Pulsation-triggered mass loss from AGB stars: the 60-day critical period}

\author{I.~McDonald\altaffilmark{1}, A.~A.~Zijlstra\altaffilmark{1}}
%%%
\altaffiltext{1}{Jodrell Bank Centre for Astrophysics, Alan Turing Building, Manchester, M13 9PL, UK; iain.mcdonald-2@jb.man.ac.uk, albert.zijlstra@manchester.ac.uk}

\begin{abstract}
Low- and intermediate-mass stars eject much of their mass during the late, red giant branch (RGB) phase of evolution. The physics of their strong stellar winds is still poorly understood. In the standard model, stellar pulsations extend the atmosphere, allowing a wind to be driven through radiation pressure on condensing dust particles. Here we investigate the onset of the wind, using nearby RGB stars drawn from the \emph{Hipparcos} catalogue. We find a sharp onset of dust production when the star first reaches a pulsation period of 60 days. This approximately co-incides with the point where the star transitions to the first overtone pulsation mode. Models of the spectral energy distributions show stellar mass-loss rate suddenly increases at this point, by a factor of $\sim$10 over the existing (chromospherically driven) wind. The dust emission is strongly correlated with both pulsation period and amplitude, indicating stellar pulsation is the main trigger for the strong mass loss, and determines the mass-loss rate. Dust emission does not strongly correlate with stellar luminosity, indicating radiation pressure on dust has little effect on the mass-loss rate. RGB stars do not normally appear to produce dust, whereas dust production by asymptotic giant branch stars appears commonplace, and is probably ubiquitous above the RGB-tip luminosity. We conclude that the strong wind begins with a step change in mass-loss rate, and is triggered by stellar pulsations. A second rapid mass-loss-rate enhancement is suggested when the star transitions to the fundamental pulsation mode, at a period of $\sim$300 days.
\end{abstract}


%%%%%%%%%%%%%%%%%%%%%%%%%%%%%%%%%%%%%%%%%%%%%%%%%%%%%%%%%%%%%%%%%%%%%%%%%%%%%%%%%%%%%%%%%%%%%%%%%%%%%%%%%

\section{Introduction}

The mechanisms triggering mass loss during the red and asymptotic giant branch (RGB/AGB) phases of stellar evolution remain unclear. Canonical theory states that stars on the RGB, and those early in their AGB evolution, have winds driven by surface magnetic fields \citep{DHA84}. Later, pulsations levitate material from the stellar surface, where it can cool and condense into dust. Radiation pressure on that dust forces it from the star, and collisional coupling with the surrounding gas means the entire circumstellar envelope can be ejected \citep[e.g.][]{WLBJ+00}. The dust reprocesses some of the radiation from the star into the mid-infrared ($\lambda \gtrsim 10$ $\mu$m), therefore the mass-loss rate can be linked to the infrared colour of the star (e.g.\ $K$--[25]). Winds driven by this combination of pulsation and radiation pressure are expected to end the evolution of all low- and intermediate-mass stars (0.8--8 M$_\odot$). But what initiates this final mass loss?

A number of studies have linked both mass-loss rate and wind velocity to stellar luminosity, suggesting that dust driving dominates the kinetics of the wind. \citet{Woitke06b} showed that momentum transferred from absorbed stellar radiation is insufficient to drive winds in oxygen-rich stars, and scattering by large dust grains is now thought to be the dominant momentum input \citep{Hoefner08,NTI+12}. This has profound implications for the early enrichment of galaxies by AGB stars: metal-poor stars cannot easily lose mass if an optically thick layer of large dust grains is required to trigger strong mass loss ($\dot{M} \gtrsim 10^{-8}$ M$_\odot$ yr$^{-1}$). Contrary to this expectation, prodigious dust production is seen in even the most metal-poor stars \citep[e.g.][]{SMM+10,MvLS+11,SML+12}.

However, the above works focus on luminous stars which are established mass losers. The inability to drive reasonable winds for metal-poor and less-evolved stars with dust-driven winds suggest that strong mass loss may be initially triggered by pulsation, rather than dust nucleation \citep{MBvLZ11,MvLS+11,MZS+14,MZS+16}. AGB stars increase in both pulsation period and amplitude as they evolve, expand and lose mass. A correlation between this increasing pulsation period and increasing mass-loss rate was shown by \citet{VW93}. Further observations and theoretical calculations have shown that a fully developed, pulsation-enhanced, dust-driven wind probably requires a star pulsating with period $P > 300$ days \citep{GWSK98,WLBJ+00,GSSP09,Uttenthaler13}, hence most studies have concentrated on the most-evolved stars, in this regime. However, substantial dust production is also seen at shorter periods. \citet{GSB+09} note a systematic increase in dust production at $P = 60$ days for Galactic stars in Baade's Window. Beyond this period, the IR excess flattens off at $K_{\rm s}-[24] \approx 1.5$ mag until $P \approx 300$ days, when there is another sharp increase (see also \citealt{BMS+15}). These sharp jumps in $K_{\rm s}-[24]$ (and their proxies, $K_{\rm s}-[22]$ and $K_{\rm s}-[25]$) indicate that pulsation period also has a strong impact on the amount of dust stars produce at $P < 300$ days.

In this paper, we examine nearby, short-period AGB stars with distances from the \emph{Hipparcos} sample \citep{vanLeeuwen07}. We study the relation between period and mass loss, and identify pulsation as a potentially important trigger of strong stellar winds.

% -------------------------------------------------------------------------------------------------------

\section{Data}

\begin{figure*}
\centerline{\includegraphics[height=0.90\textwidth,angle=-90]{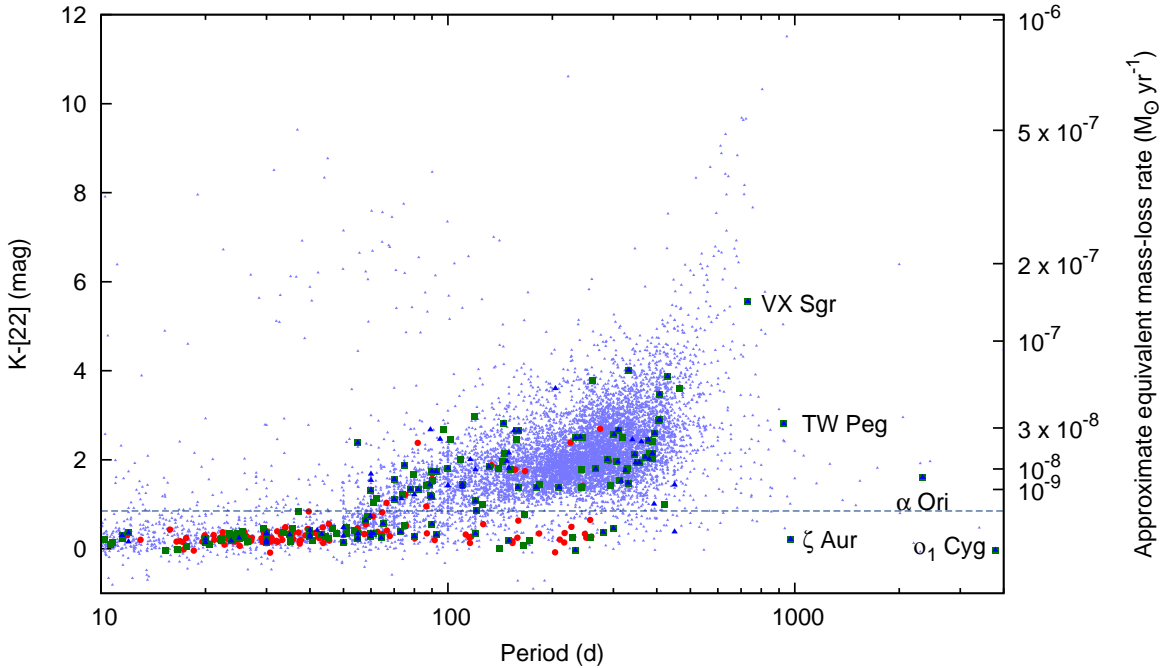}}
\caption{Relation between pulsation period and infrared excess ($K_{\rm s}$--[22]) diagram. Period data is taken from \citet{TBK+09} (red circles), VSX (green squares) and GCVS (blue triangles). Smaller, light blue triangles show GCVS stars not detected by \emph{Hipparcos} but with $K_{\rm s} < 9$ mag). The horizontal line shows our criterion for substantial dust excess.}
%Point size scales with absolute $K_{\rm s}$-band magnitude. 
\label{PXSFig}
\end{figure*}

\begin{figure*}
\centerline{\includegraphics[height=0.45\textwidth,angle=-90]{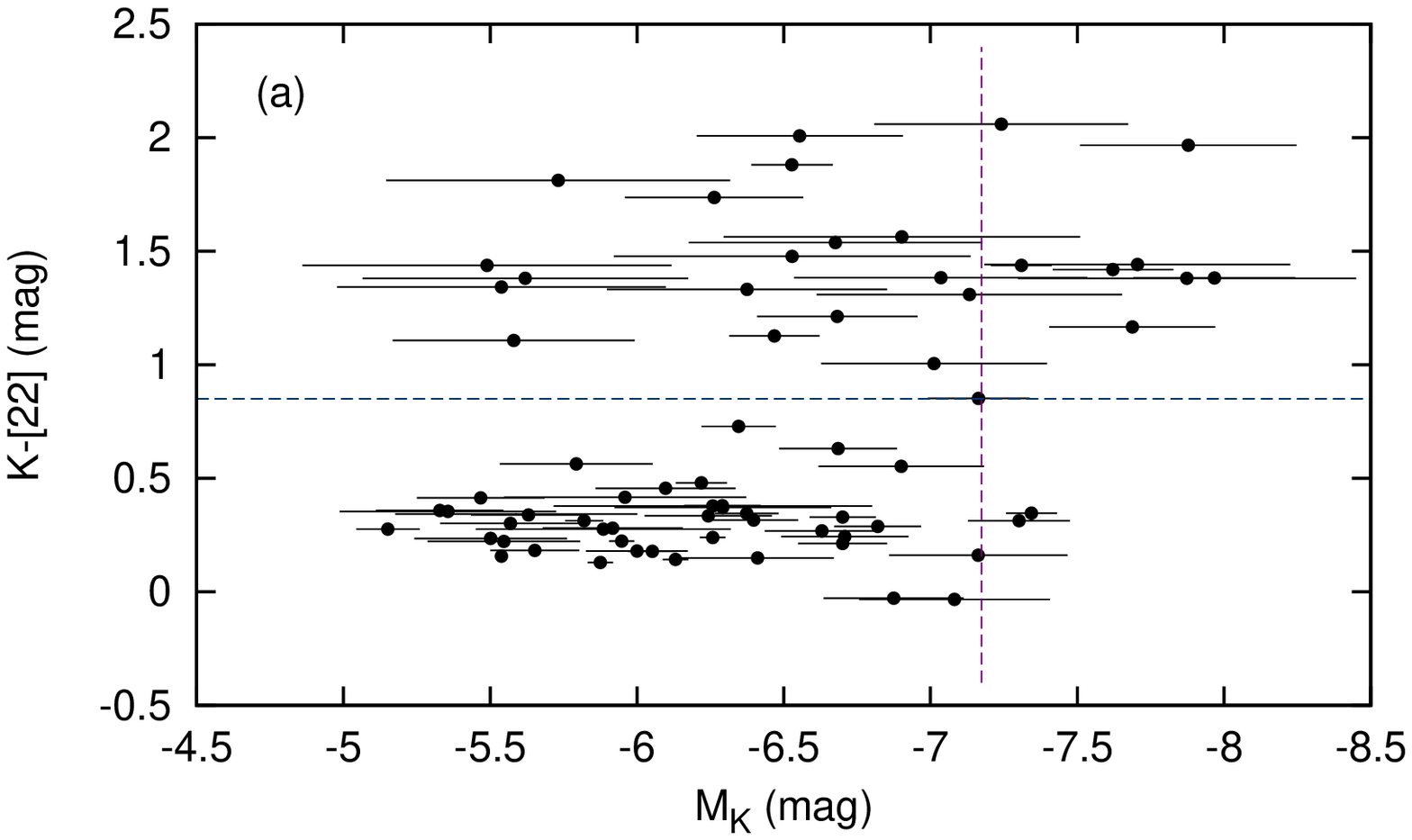}
            \includegraphics[height=0.45\textwidth,angle=-90]{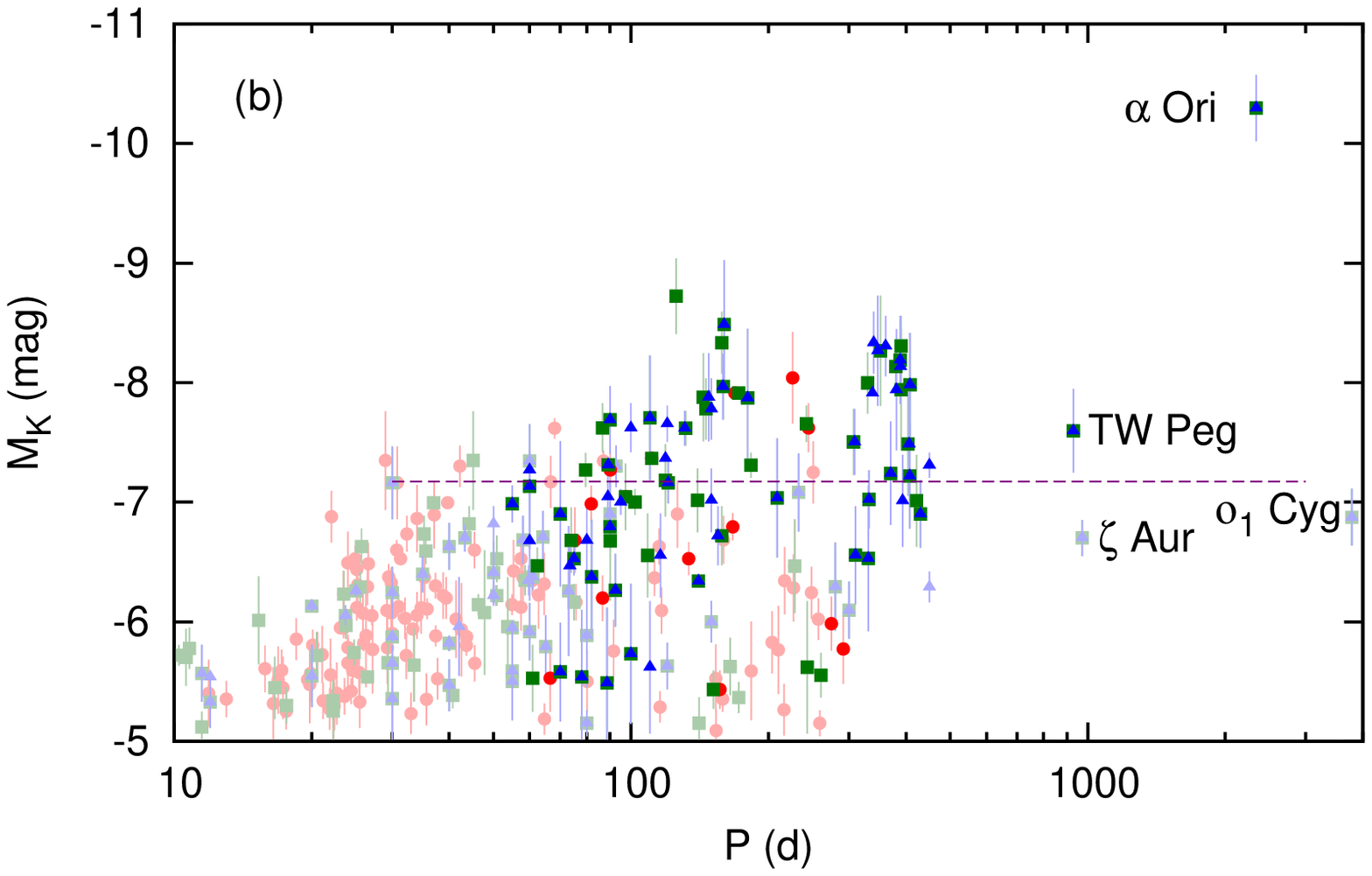}}
\centerline{\includegraphics[height=0.45\textwidth,angle=-90]{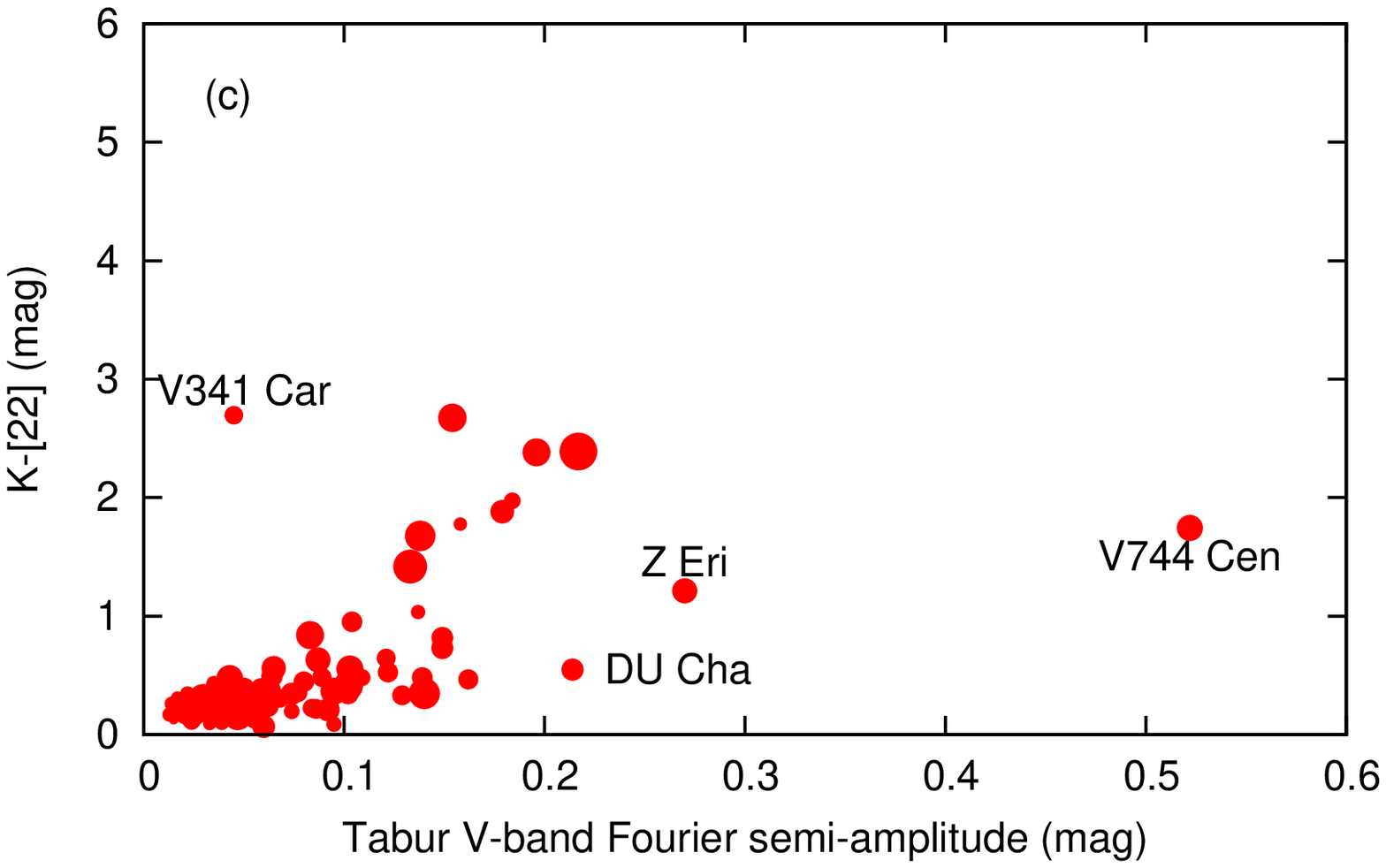}
            \includegraphics[height=0.45\textwidth,angle=-90]{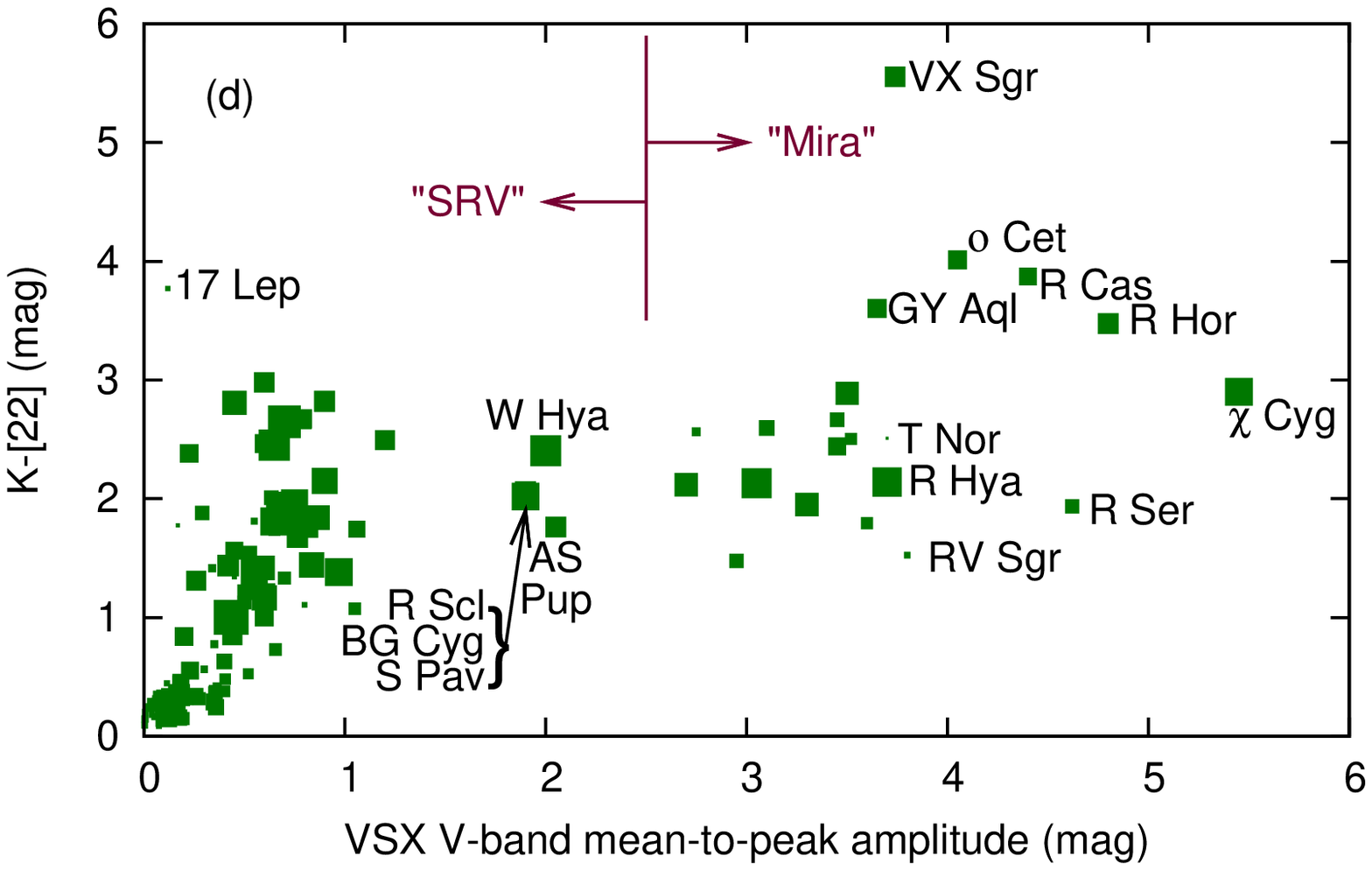}}
\centerline{\includegraphics[height=0.45\textwidth,angle=-90]{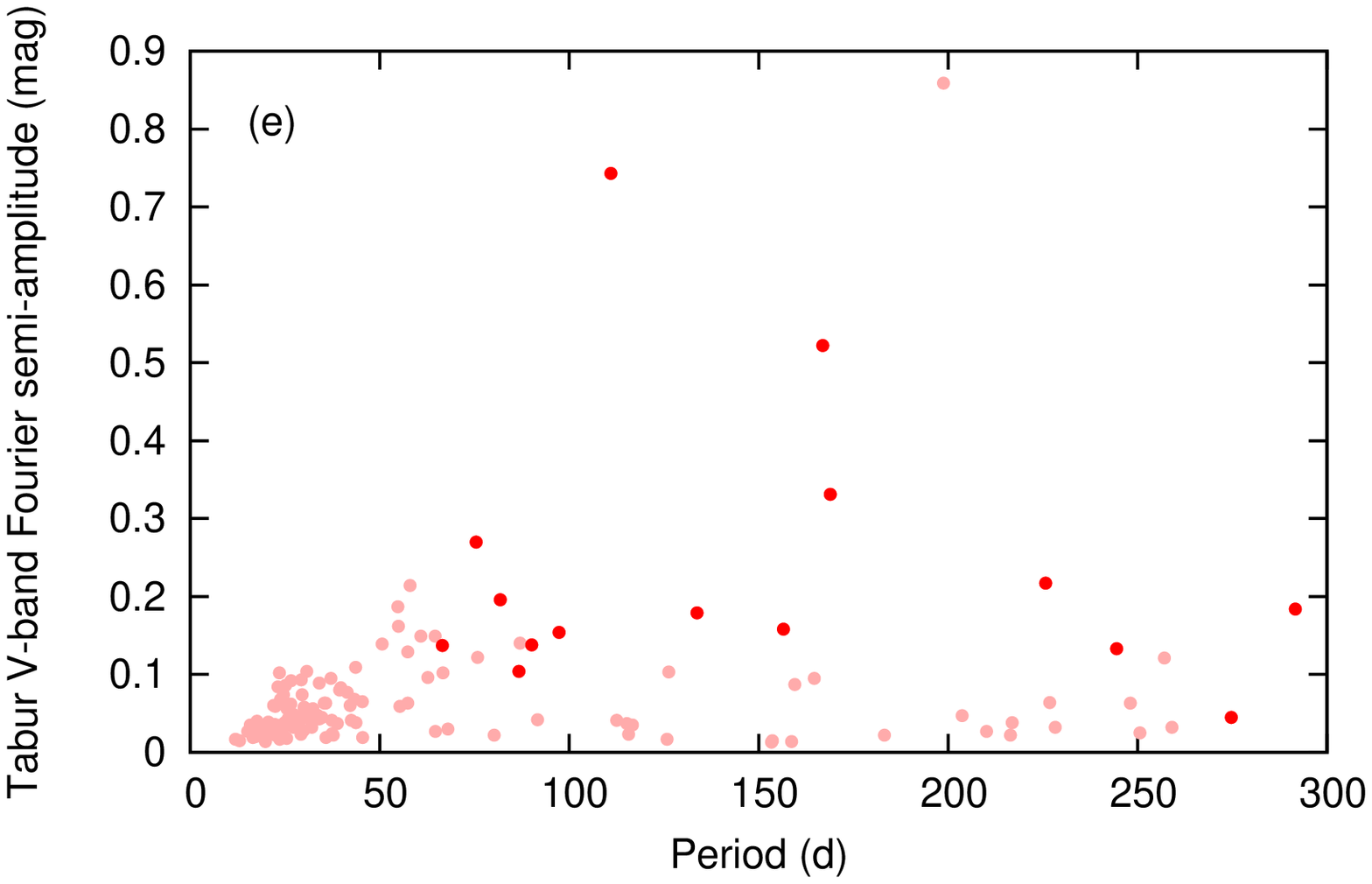}
            \includegraphics[height=0.45\textwidth,angle=-90]{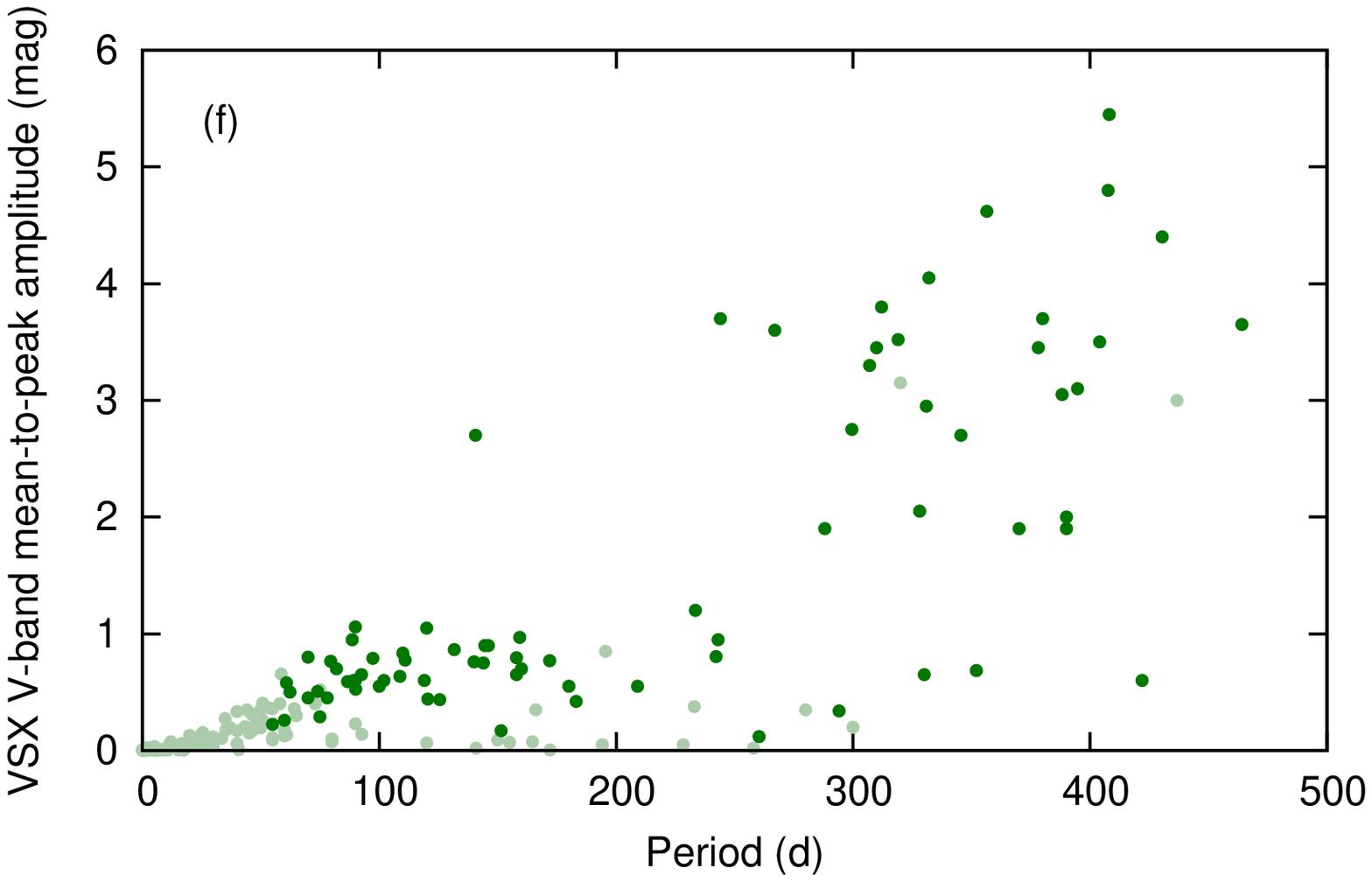}}
\caption{Relations between circumstellar dust production ($K$--[22]), and pulsation properties and luminosity. Colours are as in Figure \ref{PXSFig}. {\it (a)} Luminosity--excess diagram, based on 2MASS $K_{\rm s}$ magnitude and \emph{Hipparcos} distance. The vertical line shows the RGB tip; the horizontal line shows our $K$--[22] $>$ 0.85 mag dust excess criterion. {\it (b)} Period--luminosity diagram for the sample. Darker points show stars with $K$--[22] $>$ 0.85 mag. {\it (c)} Amplitude--excess diagrams, based on Fourier-decomposed amplitudes from \citet{TBK+09} and {\it (d)} and mean-to-peak amplitudes from GCVS. {\it (e) \& (f)} Period--amplitude diagrams, as second row. Light-coloured stars have little dust production ($K-[22] < 0.85$ mag).} % {\it Bottom right:} Temperature--excess diagram.}
\label{XSFig}
\end{figure*}

Our dataset comprises of a sample of nearby giant stars. Initially, 560 stars with parallax distances of $d < 300$ pc were selected from the \emph{Hipparcos} catalogue \citep{vanLeeuwen07}. Most of these stars have luminosities ($L$) and temperatures ($T_{\rm eff}$) determined by \citet{MZB12}, and a cut was made to only include giant stars with $L > 680$ L$_\odot$ and $T_{\rm eff} < 5000$ K. $L > 680$ L$_\odot$ represents the luminosity where substantial infrared excess due to circumstellar dust production is first seen. Some bright stars lacked unsaturated photometry in the surveys used in \citet{MZB12}, so were not included in this initial list. We have manually added missing bright ($M_{\rm Ks,\ 2MASS} < 0$; \citealt{CSvD+03}) stars from the \emph{Hipparcos} catalogue. This cutoff approximates our criteria of $L > 680$ L$_\odot$ and $T_{\rm eff} < 5000$ K, but may also lead to the inclusion of some fainter stars. The only star likely to meet the selection criteria and be within 300 pc, but not included in \emph{Hipparcos}, is the optically enshrouded star CW Leo. With its addition, we obtain a near-complete sample of giant stars with $L > 680$ L$_\odot$ within 300 pc. Some stars may still be erroneously included or excluded depending on their $K_{\rm s}$-band bolometric corrections and parallax uncertainties, which affect their derived distances and luminosities.

Infrared photometry for each star has been drawn from the 2MASS and {\it WISE} surveys \citep{CWC+13}, allowing us to construct $K_{\rm s}-[22]$ colours which are representative of the dust column density in front of the star. For bright stars, the {\it WISE} photometry displays inaccuracies. Sources with 25-$\mu$m {\it IRAS} fluxes \citep{BHW88} of $F_{25} > 115$ Jy were found to strongly disagree with {\it WISE} results. For these stars, we have interpolated between the 12-$\mu$m and 25-$\mu$m {\it IRAS} fluxes in log($F$)--log($\lambda$) to provide a 22-$\mu$m flux. For HIP 113249, {\it IRAS} 12-$\mu$m photometry was unavailable. For this star we simply use the 25-$\mu$m flux to represent the 22-$\mu$m flux.

Pulsation data has been obtained from three sources. Most optically bright stars have $V$-band Fourier-decomposed periods and amplitudes listed in \citet{TBK+09}. The period with the highest amplitude was chosen as representative for the star. Stars in the \citet{TBK+09} sample are limited to $V \lesssim 8$ mag and periods of $P < 300$ days, thus miss the reddened, longer-period pulsators. Additional $V$-band periods and peak-to-peak amplitudes from VSX \citep{Watson06}, and periods from GCVS \citep{SDZ+06} were adopted. The VSX amplitudes are generally much larger than those of \citet{TBK+09} due to the addition of harmonics and variability on timescales of $P > 300$ days. VSX and GCVS lightcurves also extend to redder objects with higher amplitude variations, as they are sensitive to optically fainter stars.

The final sample contains 519 stars with an \emph{Hipparcos} distance, at least one measure of variability and a $K_{\rm s}-[22]$ colour. We also retain separately a list of stars in the GCVS without \emph{Hipparcos} distances, but with $K_{\rm s}-[22]$ colours. We limit these to stars with $K_{\rm s}<9$ mag, to avoid sensitivity limits in the [22] data.

Most stars should be within a factor of two of the solar metallicity \citep{TC05}. Because of this, the sample is almost entirely oxygen-rich. There are only two known carbon stars in the sample: U Hya and RT Cap. No known S stars are included.

In order to approximately convert $K_{\rm s}$--[22] colour to mass-loss rate, we use the \citet{SZK94} version of the \citet{SK92} dust model, assuming efficient condensation into silicate dust and a silicon fraction of 10 per million atoms. Effective dust-to-gas collisional coupling and a 10 km s$^{-1}$ outflow velocity were assumed. This $K_{\rm s}$--[22] colour to mass-loss rate conversion compares well to the $K_{\rm s}$--[25] relation seen in nearby stars \citep[e.g.][]{vanLoon07}. However, uncertainties and differences in stellar luminosity, grain properties, condensation efficiency and outflow velocity limit the accuracy of this conversion to a factor of a few.

% -------------------------------------------------------------------------------------------------------

\section{Results}

Figure \ref{PXSFig} shows that dust-production rate clearly increases sharply as soon as stars reach the critical 60-day period identified by \citet{GSB+09}. After this initial jump, the amount of dust excess and the fraction of stars with dust excess both continue to increase to $P \approx 120$ days. Following this, dust excess may plateau between $120 \lesssim P \lesssim 300$ days, before increasing again at longer periods. In the \emph{Hipparcos} 519-star sample, stars with $P > 500$ days or $K_{\rm s}-[22] > 4$ mag are supergiants ($\alpha$ Ori) or known binary stars. At high mass-loss rates, the $K_{\rm s}-[22]$ colour will saturate as the SED peak becomes longer than 22 $\mu$m. However, only the most extreme stars listed here ($K_{\rm s}-[22] \gg 4$) should fall into that high-mass-loss-rate regime.

A significant number of stars retain colours $K_{\rm s}-[22] < 0.85$ mag but have periods $P \gg 60$ days. The period--luminosity diagram (Figure \ref{XSFig}(b)) shows that these stars are found predominantly below the red giant branch (RGB) tip, while stars with infrared excess are distributed nearly evenly across the RGB tip. Above the RGB tip, 93\% of stars have excess. The remaining 7\% mostly have $P < 60$ days and are probably RGB stars with uncertain distances. This suggests that the dust-producing stars are mostly or entirely AGB stars, and that the RGB stars produce little or no dust.

The relation between pulsation amplitude and $K_{\rm s}$--[22] colour is clearer (Figure \ref{XSFig}(c) \& (d)). The pulsation amplitudes of both the \citet{TBK+09} and VSX samples correlate very strongly with $K_{\rm s}$--[22] colour for low amplitude variables. However, in the VSX data, this relationship breaks down at $\delta V \approx 1$ mag. This is a smaller amplitude than the classical Mira--semi-regular-variable (SRV) split at $\delta V = 2.5$ mag. Stars in the region of $\delta V = 1$--2.5 mag (named on Figure \ref{XSFig}(d)) still tend to have very regular variability, hence belong to the paradoxically termed SRa class of ``regular semi-regular'' variables. The correlation between amplitude and $K_{\rm s}$--[22] colour picks up again at $\delta V \approx 3.5$ mag.

This gap between $\delta V \approx 1$ and 3.5 mag where the amplitude--colour relation breaks down can be explained by examining the period--amplitude diagrams (Figure \ref{XSFig} (e) \& (f)). These diagrams show a correlation between period and amplitude up to $P \approx 100$ days and. In the VSX data, stars with $P \approx$ 100 days typically have $\delta V \sim 1$ mag. In the regime $100 \lesssim P \lesssim 300$ days, the correlation breaks down: few stars are found with $\delta V > 1$ mag. The same is true of the \citet{TBK+09} data, although the Fourier-based amplitude is correspondingly lower ($\delta V \sim 0.3$ mag). In the VSX data, the correlation restarts at $P \sim 300$ days and $\delta V \sim 3$ mag. This hiatus between 100 and 300 days is commensurate with the rapid transition of many stars from being semi-regular variables on the first overtone sequence of the period--luminosity diagram to being `Mira' variables on the fundamental pulsation sequence \citep[e.g.][]{BMS+15,Wood15}.

For stars which do produce dust ($K_{\rm s}-[22] > 0.85$ mag), one would expect a dust-driven wind to produce a correlation between dust-mass-loss rate and luminosity, but none is seen. Figure \ref{XSFig}(a) shows a large amount of scatter among the dust-producing stars on both sides of the RGB tip. These stars are not expected to differ widely in metallicity (95\% of stars should be within --0.3 $\lesssim$ [Fe/H] $\lesssim$ +0.3 dex; \citealt{TC05}). Nor should they vary much in C/O (C/O $\approx$ 0.4 for most stars, up to $\approx$ 0.9 for a few of the most-evolved stars currently experiencing dredge-up, plus the two carbon stars). The primary difference among stars at given luminosity is therefore expected to be (current) mass. This may vary by an order of magnitude (0.53--8 M$\odot$). However, the initial mass function of stars, the relatively low star-formation rate of the solar neighbourhood in recent cosmological times, and the rapid increase in mass-loss rate towards the end of a star's life should dramatically bias the mass distribution to the range of $\sim$0.7--1.4 M$_\odot$ \citep[e.g.][]{BR92,AB09}. The dust-producing stars shown in Figure \ref{XSFig}(a) cover a factor of $\sim$10 in luminosity yet, despite the expected homogeneity of the stellar parameters, no correlation is seen between luminosity and infrared dust excess.

The various plots in Figure \ref{XSFig} show that both bright and faint stars produce dust, therefore luminosity is not the primary driver of dust production in most local AGB stars, {and stellar mass-loss rate cannot be easily defined in terms of luminosity}. Pulsation amplitude and dust production are correlated, but both the small- and large-amplitude pulsators still produce dust. Of all the related parameters (luminosity, period and amplitude), the clearest onset of dust production comes with the change in pulsation period (Figure \ref{PXSFig}).

% -------------------------------------------------------------------------------------------------------

\section{Discussion}

The onset of strong dust production at a 60-day pulsation period in AGB stars, already found in Baade's Window and (tentatively) towards the LMC, also exists in local stars. These three very different populations have different star-formation histories (i.e.\ different stellar masses) and different metallicities, and the consistency of this 60-day period among them strongly suggests that it is the trigger of stellar dust production. The question remains as to whether this is a true step function in stellar mass-loss rate, or whether it is simply that dust begins to condense in an already-established wind.

Few homogeneously derived gas mass-loss rates exist for stars with such short periods. Some constraint comes from \citet{Groenewegen12} and \citet{MZS+15}, which examine two very similar stars close to the 60-day boundary: VY Leo and EU Del, respectively. Both are barely producing dust ($K_{\rm s}-[22] = 0.465$ and 0.729 mag, respectively). Their gas mass-loss rates are a few $\times$10$^{-9}$ and a few $\times$10$^{-8}$ M$_\odot$ yr$^{-1}$, respectively. Both dust and gas mass-loss rates are difficult to determine with absolute accuracy, but the similarities in their outflow velocities indicate that they should have good relative accuracy. Accounting for the likely metal-paucity of EU Del ([Fe/H] $\approx$ --0.27 dex), these are commensurate with our adopted mass-loss-rate to $K_{\rm s}-[22]$ colour conversion, {\it and indicate that the dust:gas ratio does not change appreciably as one approaches the 60-day transition. This suggests} dust condensation is roughly as efficient for stars below and above our $K_{\rm s}-[22] = 0.85$ mag cutoff, {although further data for stars across the 60-day transition will be needed to show this conclusively}.
%They also have similar wind speeds (12 and 9.5 km s$^{-1}$, respectively).

The significance of the 60-day period remains undetermined. \citet{MDB+13} fit a relationship between pulsation period and amplitude. For this relationship, pulsation acceleration approaches the surface gravity of the star at a period of $\sim$60 days. This would allow the ballistic motion of the pulsation to eject the outer envelope of the star, with the mass-loss rate being set by the density of the unbound region. The sharp increase at 60 days would therefore correspond to the unbound region migrating inwards to reach a density jump in the stellar atmosphere. However, we note that the \citet{MDB+13} analysis ignores stellar temperature variations, whereas we know that the spectral types of these stars change considerably \citep{SDZ+06}.

We have shown that stars can start to produce dust at 60 days, but not all do. Their distribution across the RGB tip indicates the dusty stars are AGB stars. However, the surface properties of AGB and RGB stars are very similar at this point. Observations in the Magellanic Clouds show that RGB stars typically populate the second and third overtone pulsations, but not the first, which is apparently exclusively reserved for AGB stars \citep{Soszynski09}. As noted above, the vast majority of evolved stars in the solar neighbourhood have present-day masses in the range $\sim$0.7--1.4 M$_\odot$, hence initial masses in the range $\sim$0.8--1.4 M$_\odot$. Periods of 50--80 days represent the evolutionary phase where AGB stars of 0.8--1.4 M$_\odot$ will transition onto the first overtone period (sequence C$^\prime$ in \citet[][figure 10]{Wood15}). It may be that pulsation on this sequence is required to initiate dust production, as noted by \citet{BMS+15}.

The slightly lower mean metallicity of the Magellanic Clouds (canonically [Fe/H] = --0.3 dex) results in a slightly different timing of the transition between pulsation modes. For a star of given mass and given luminosity (near the RGB tip), stellar radius of a star at [Fe/H] = 0 dex will be $\approx$15\% larger than one at [Fe/H] = --0.3 dex due to the increased atmospheric metal-line and molecular opacity \citep[e.g.][]{DCJ+08}. This implies a density ($\rho$) decrease of $\sim$44\%. Since $P \propto \sqrt{\rho}$ \citep[e.g.][]{CS15}, the pulsation period should be $\approx$22\% or $\sim$13 days longer for the [Fe/H] = 0 dex star compared to the [Fe/H] = --0.3 dex star. While pulsation period clearly appears to be a major factor in driving strong stellar mass loss, other factors (radius, temperature, etc.) may also be important. These may be traced by repeating our present study in environments of differing metallicity.

The correlations we identify with pulsation period, amplitude and sequence indicate that pulsation triggers the dusty wind of evolved AGB stars. The lack of correlation between dust excess and luminosity shows that dust driving is not an important factor at this stage, although it may become important later. We therefore reach the following conclusions:
\begin{enumerate}
\item Strong mass loss does not normally begin until the pulsation period exceeds $\sim$60 days.
\item Most AGB stars with $P \gtrsim 120$ days produce copious dust.
\item Strong mass loss starts here for AGB stars, but not typically for RGB stars.
\item Observations of VY Leo and EU Del indicate that this appears to be an increase in mass-loss rate, not just dust-production rate. While not conclusively shown, it may be linked to pulsations providing a ballistic trajectory to the outermost atmosphere which allows it to escape from the star without substantial radiation pressure on dust.
\item Pulsation period and amplitude may together define the mass-loss rate during this early dust-producing phase. A survey of circumstellar CO lines for stars in this regime is suggested, in order that a relation can be defined between pulsation properties, and stellar mass-loss rates and terminal wind velocities.
\item The data are still consistent with the previously published increase in mass-loss rate at $P \sim 300$ days. This would correspond to the stars moving from the first overtone to the fundamental mode. However, relating pulsation period to mass-loss rate likely requires a step function to accommodate this transition.
\end{enumerate}

% -------------------------------------------------------------------------------------------------------

%\section{Conclusions}

%\vspace{5 mm}

%{\bf \color{red} XXX Acknowledgements}

%\bibliographystyle{apj}
%\bibliography{references}

\begin{thebibliography}{38}
\expandafter\ifx\csname natexlab\endcsname\relax\def\natexlab#1{#1}\fi

\bibitem[{{Aumer} \& {Binney}(2009)}]{AB09}
{Aumer}, M., \& {Binney}, J.~J. 2009, MNRAS, 397, 1286

\bibitem[{{Basu} \& {Rana}(1992)}]{BR92}
{Basu}, S., \& {Rana}, N.~C. 1992, ApJ, 393, 373

\bibitem[{{Beichmann} {et~al.}(1988){Beichmann}, {Helou}, \& {Walker}}]{BHW88}
{Beichmann}, C.~A., {Helou}, G., \& {Walker}, D.~W. 1988, {Infrared
  astronomical satellite (IRAS). Catalogs and atlases} (NASA RP (Reference
  Publication). NASA, Washington.)

\bibitem[{{Boyer} {et~al.}(2015){Boyer}, {McDonald}, {Srinivasan}, {Zijlstra},
  {van Loon}, {Olsen}, \& {Sonneborn}}]{BMS+15}
{Boyer}, M.~L., {McDonald}, I., {Srinivasan}, S., {Zijlstra}, A., {van Loon},
  J.~T., {Olsen}, K.~A.~G., \& {Sonneborn}, G. 2015, ApJ, 810, 116

\bibitem[{{Catelan} \& {Smith}(2015)}]{CS15}
{Catelan}, M., \& {Smith}, H.~A. 2015, {Pulsating Stars}

\bibitem[{{Cutri} {et~al.}(2003){Cutri}, {Skrutskie}, {van Dyk}, {Beichman},
  {Carpenter}, {Chester}, {Cambr{\'e}sy}, {Evans}, {Fowler}, {Gizis}, {Howard},
  {Huchra}, \& {et al.}}]{CSvD+03}
{Cutri}, R.~M., {et~al.} 2003, {2MASS All Sky Catalog of point sources.} (The
  IRSA 2MASS All-Sky Point Source Catalog, NASA/IPAC Infrared Science Archive.)

\bibitem[{{Cutri} {et~al.}(2013){Cutri}, {Wright}, {Conrow}, {Fowler},
  {Eisenhardt}, {Grillmair}, {Kirkpatrick}, {Masci}, {McCallon}, {Wheelock},
  {Fajardo-Acosta}, {Yan}, {Benford}, {Harbut}, {Jarrett}, {Lake}, {Leisawitz},
  {Ressler}, {Stanford}, {Tsai}, {Liu}, {Helou}, {Mainzer}, {Gettings},
  {Gonzalez}, {Hoffman}, {Marsh}, {Padgett}, {Skrutskie}, {Beck}, {Papin}, \&
  {Wittman}}]{CWC+13}
---. 2013, {Explanatory Supplement to the AllWISE Data Release Products}, Tech.
  rep.

\bibitem[{{Dotter} {et~al.}(2008){Dotter}, {Chaboyer}, {Jevremovi{\'c}},
  {Kostov}, {Baron}, \& {Ferguson}}]{DCJ+08}
{Dotter}, A., {Chaboyer}, B., {Jevremovi{\'c}}, D., {Kostov}, V., {Baron}, E.,
  \& {Ferguson}, J.~W. 2008, ApJS, 178, 89

\bibitem[{{Dupree} {et~al.}(1984){Dupree}, {Hartmann}, \& {Avrett}}]{DHA84}
{Dupree}, A.~K., {Hartmann}, L., \& {Avrett}, E.~H. 1984, ApJ, 281, L37

\bibitem[{{Glass} {et~al.}(2009){Glass}, {Schultheis}, {Blommaert}, {Sahai},
  {Stute}, \& {Uttenthaler}}]{GSB+09}
{Glass}, I.~S., {Schultheis}, M., {Blommaert}, J.~A.~D.~L., {Sahai}, R.,
  {Stute}, M., \& {Uttenthaler}, S. 2009, MNRAS, 395, L11

\bibitem[{{Groenewegen}(2012)}]{Groenewegen12}
{Groenewegen}, M.~A.~T. 2012, A\&A, 540, A32

\bibitem[{{Groenewegen} {et~al.}(2009){Groenewegen}, {Sloan}, {Soszy{\'n}ski},
  \& {Petersen}}]{GSSP09}
{Groenewegen}, M.~A.~T., {Sloan}, G.~C., {Soszy{\'n}ski}, I., \& {Petersen},
  E.~A. 2009, A\&A, 506, 1277

\bibitem[{{Groenewegen} {et~al.}(1998){Groenewegen}, {Whitelock}, {Smith}, \&
  {Kerschbaum}}]{GWSK98}
{Groenewegen}, M.~A.~T., {Whitelock}, P.~A., {Smith}, C.~H., \& {Kerschbaum},
  F. 1998, MNRAS, 293, 18

\bibitem[{{H{\"o}fner}(2008)}]{Hoefner08}
{H{\"o}fner}, S. 2008, A\&A, 491, L1

\bibitem[{{McDonald} {et~al.}(2011{\natexlab{a}}){McDonald}, {Boyer}, {van
  Loon}, \& {Zijlstra}}]{MBvLZ11}
{McDonald}, I., {Boyer}, M.~L., {van Loon}, J.~T., \& {Zijlstra}, A.~A.
  2011{\natexlab{a}}, ApJ, 730, 71

\bibitem[{{McDonald} {et~al.}(2012){McDonald}, {Zijlstra}, \& {Boyer}}]{MZB12}
{McDonald}, I., {Zijlstra}, A.~A., \& {Boyer}, M.~L. 2012, MNRAS, 427, 343

\bibitem[{{McDonald} {et~al.}(2014){McDonald}, {Zijlstra}, {Sloan}, {Kerins},
  {Lagadec}, \& {Minniti}}]{MZS+14}
{McDonald}, I., {Zijlstra}, A.~A., {Sloan}, G.~C., {Kerins}, E., {Lagadec}, E.,
  \& {Minniti}, D. 2014, MNRAS, 439, 2618

\bibitem[{{McDonald} {et~al.}(2015){McDonald}, {Zijlstra}, {Sloan}, {Lagadec},
  {Johnson}, {Uttenthaler}, {Jones}, \& {Smith}}]{MZS+15}
{McDonald}, I., {Zijlstra}, A.~A., {Sloan}, G.~C., {Lagadec}, E., {Johnson},
  C.~I., {Uttenthaler}, S., {Jones}, O.~C., \& {Smith}, C.~L. 2015, arXiv:1512.04695

\bibitem[{{McDonald} {et~al.}(2016){McDonald}, {Zijlstra}, {Sloan}, {Lagadec},
  {Johnson}, {Uttenthaler}, {Jones}, \& {Smith}}]{MZS+16}
---. 2016, MNRAS, 456, 4542

\bibitem[{{McDonald} {et~al.}(2011{\natexlab{b}}){McDonald}, {van Loon},
  {Sloan}, {Dupree}, {Zijlstra}, {Boyer}, {Gehrz}, {Evans}, {Woodward}, \&
  {Johnson}}]{MvLS+11}
{McDonald}, I., {et~al.} 2011{\natexlab{b}}, MNRAS, 417, 20

\bibitem[{{Mosser} {et~al.}(2013){Mosser}, {Dziembowski}, {Belkacem}, {Goupil},
  {Michel}, {Samadi}, {Soszy{\'n}ski}, {Vrard}, {Elsworth}, {Hekker}, \&
  {Mathur}}]{MDB+13}
{Mosser}, B., {et~al.} 2013, A\&A, 559, A137

\bibitem[{{Norris} {et~al.}(2012){Norris}, {Tuthill}, {Ireland}, {Lacour},
  {Zijlstra}, {Lykou}, {Evans}, {Stewart}, \& {Bedding}}]{NTI+12}
{Norris}, B.~R.~M., {et~al.} 2012, Nature, 484, 220

\bibitem[{{Samus} {et~al.}(2006){Samus}, {Durlevich}, {Zharova}, {Kazarovets},
  {Kireeva}, {Pastukhova}, {Williams}, \& {Hazen}}]{SDZ+06}
{Samus}, N.~N., {Durlevich}, O.~V., {Zharova}, A.~V., {Kazarovets}, E.~V.,
  {Kireeva}, N.~N., {Pastukhova}, E.~N., {Williams}, D.~B., \& {Hazen}, M.~L.
  2006, Astronomy Letters, 32, 263

\bibitem[{{Siebenmorgen} \& {Kruegel}(1992)}]{SK92}
{Siebenmorgen}, R., \& {Kruegel}, E. 1992, A\&A, 259, 614

\bibitem[{{Siebenmorgen} {et~al.}(1994){Siebenmorgen}, {Zijlstra}, \&
  {Kr{\"u}gel}}]{SZK94}
{Siebenmorgen}, R., {Zijlstra}, A.~A., \& {Kr{\"u}gel}, E. 1994, MNRAS, 271,
  449

\bibitem[{{Sloan} {et~al.}(2010){Sloan}, {Matsunaga}, {Matsuura}, {Zijlstra},
  {Kraemer}, {Wood}, {Nieusma}, {Bernard-Salas}, {Devost}, \& {Houck}}]{SMM+10}
{Sloan}, G.~C., {et~al.} 2010, ApJ, 719, 1274

\bibitem[{{Sloan} {et~al.}(2012){Sloan}, {Matsuura}, {Lagadec}, {van Loon},
  {Kraemer}, {McDonald}, {Groenewegen}, {Wood}, {Bernard-Salas}, \&
  {Zijlstra}}]{SML+12}
---. 2012, ApJ, 752, 140

\bibitem[{{Soszy{\'n}ski}(2009)}]{Soszynski09}
{Soszy{\'n}ski}, I. 2009, in IAU Symposium, Vol. 256, IAU Symposium, ed. J.~T.
  {van Loon} \& J.~M. {Oliveira}, 30--35

\bibitem[{{Tabur} {et~al.}(2009){Tabur}, {Bedding}, {Kiss}, {Moon}, {Szeidl},
  \& {Kjeldsen}}]{TBK+09}
{Tabur}, V., {Bedding}, T.~R., {Kiss}, L.~L., {Moon}, T.~T., {Szeidl}, B., \&
  {Kjeldsen}, H. 2009, MNRAS, 400, 1945

\bibitem[{{Taylor} \& {Croxall}(2005)}]{TC05}
{Taylor}, B.~J., \& {Croxall}, K. 2005, MNRAS, 357, 967

\bibitem[{{Uttenthaler}(2013)}]{Uttenthaler13}
{Uttenthaler}, S. 2013, A\&A, 556, A38

\bibitem[{{van Leeuwen}(2007)}]{vanLeeuwen07}
{van Leeuwen}, F. 2007, A\&A, 474, 653

\bibitem[{{van Loon}(2007)}]{vanLoon07}
{van Loon}, J.~T. 2007, in Astronomical Society of the Pacific Conference
  Series, Vol. 378, Why Galaxies Care About AGB Stars: Their Importance as
  Actors and Probes, ed. F.~{Kerschbaum}, C.~{Charbonnel}, \& R.~F. {Wing}, 227

\bibitem[{{Vassiliadis} \& {Wood}(1993)}]{VW93}
{Vassiliadis}, E., \& {Wood}, P.~R. 1993, ApJ, 413, 641

\bibitem[{{Watson}(2006)}]{Watson06}
{Watson}, C.~L. 2006, Society for Astronomical Sciences Annual Symposium, 25,
  47

\bibitem[{{Winters} {et~al.}(2000){Winters}, {le Bertre}, {Jeong}, {Helling},
  \& {Sedlmayr}}]{WLBJ+00}
{Winters}, J.~M., {le Bertre}, T., {Jeong}, K.~S., {Helling}, C., \&
  {Sedlmayr}, E. 2000, A\&A, 361, 641

\bibitem[{{Woitke}(2006)}]{Woitke06b}
{Woitke}, P. 2006, A\&A, 460, L9

\bibitem[{{Wood}(2015)}]{Wood15}
{Wood}, P.~R. 2015, MNRAS, 448, 3829

\end{thebibliography}

\end{document}